\documentclass[11pt]{article}
\newcommand{\beq}{\begin{equation}}
\newcommand{\eeq}[1]{\label{#1}\end{equation}}
\newcommand{\bea}{\begin{eqnarray}}
\newcommand{\eea}[1]{\label{#1}\end{eqnarray}}
\begin{document}
\baselineskip 18pt
\begin{titlepage}
\hfill hep-th/0402038
\vspace{20pt}

\begin{center}
{\large\bf BOUNDS ON GENERIC HIGH-ENERGY PHYSICS MODIFICATIONS TO THE 
PRIMORDIAL POWER SPECTRUM FROM BACK-REACTION ON THE METRIC}
\end{center}

\vspace{6pt}

\begin{center}
{\large M. Porrati} \vspace{20pt}

{\em Department of Physics\\ New York University\\ 4 Washington Place\\ 
New York NY 10003, USA}

\end{center}

\vspace{12pt}

\begin{center}
{\bf Abstract}
\end{center}
\begin{quotation}\noindent
Modifications to the primordial power spectrum of inflationary density 
perturbations have been studied recently using a boundary effective field 
theory approach. In the approximation of a fluctuating quantum field on a 
fixed background, the generic effect of new physics is encoded in parameters
of order $H/M$. Here, we point out that the back-reaction on the metric can be
neglected only when these parameters obey certain bounds that may put 
them beyond the reach of observation.
\end{quotation}
\vfill
 \hrule width 5.cm
\vskip 2.mm
{\small \noindent e-mail: massimo.porrati@nyu.edu}
\end{titlepage}
\section{Introduction}
The possibility of observing very high-energy, ``trans-planckian'' 
physics in the cosmic 
microwave background radiation, thanks to the enormous stretch in proper
distance due to inflation is one of the most exciting possibility for probing
string theory, or any other model for quantum gravity. As such, it has received
enormous attention, once the possibility was raised that these
effects can be as large as $H/M$, with $H$ the Hubble parameter during 
inflation, and $M$ the scale of new physics (e.g. the string scale).
A partial, derivative list of references is~\cite{all}.

Due to our ignorance of the ultimate theory governing high-energy physics,
the most natural, model-independent approach to study any modifications to the
primordial power spectrum is effective field theory~\cite{kkls,bch}. Using
an effective field theory approach, the authors of~\cite{kkls} concluded that
the signature of any trans-planckian modification of the standard inflationary
power spectrum is $O(H^2/M^2)$, well beyond the reach of observation even in 
the most favorable scenario ($H\sim 10^{14}\,\mbox{GeV}, M\sim 10^{16}\,
\mbox{Gev}$).

This conclusion has been recently criticized in~\cite{sspds}. There, it was 
pointed out that the effect of 
high-energy physics manifests in two ways. First, through the appearance
of irrelevant operators in the four-dimensional
local field theory describing physics below the
cutoff. The most relevant of these effects, studied in~\cite{kkls}, 
are parametrized by local operators of dimension six, so, they are 
naturally of order $H^2/M^2$.
Second, through a change in the initial conditions at some (arbitrary)
early time $t_0$. This change of the initial quantum state can also be 
described in the  effective field theory language, 
by adding appropriate boundary
terms at the initial-time hypersurface. In~\cite{sspds}, it was shown that 
the most relevant change in intital conditions is  
described in the three-dimensional boundary Lagrangian 
by operators of dimension 
four. They do generate effects of order $H/M$, generically. 

Reference~\cite{sspds} does not take into account all effects due to the
back-reaction of the modified stress-energy tensor on the metric. 
Specifically, any change in the boundary conditions of the effective field 
theory generates {\em finite, regularization-independent} 
modifications to expectation value of the matter stress-energy tensor. These
modifications can become large near the (space-like) boundary hypersurface.
By requiring that the back-reaction remain under control, we shall get new 
bounds on the size of the parameters encoding new physics. These bounds 
depend on some mild assumptions on the inflationary potential.
Ranging from the weakest, most generic bound to the strongest, least
generic one, we get:
\bea
\beta&<& {4\pi\over GM^2} \left({H\over M}\right)^2, \label{1} \\ \beta&<&
\epsilon {8\pi\over 5 GM^2} \left({H\over M}\right)^3, \label{2} \\ 
\beta&<& \epsilon \eta {8\pi \over 15 GM^2}\left({H\over M}\right)^4 , 
\eea{3}
Here $\beta$ is the dimensionless quantity parametrizing the effects of
high-energy physics, while $\epsilon$ and $\eta$ are standard quantities
parametrizing slow-roll inflation. 
They will all be defined in Section 3~\footnote{Bounds on $\beta$ were also
derived on somewhat different physical grounds in~\cite{st}.}
   
In this note, we re-derive for completeness the result of~\cite{sspds}, in a 
formalism appropriate to our purpose, and then we proceed to derive the
bounds in Eq.~(\ref{3}). Back-reaction in inflation
was also considered within a different approach in~\cite{g}.
\section{Boundary Interactions and Their Effect}
We follow the formalism developed in ref~\cite{sspds}, where changes in the
inflationary vacuum were parametrized by higher-dimension operators in the
boundary conditions imposed on the inflaton (or any other fluctuating field)
at an arbitrary ``initial time'' hypersurface. 
 
We use the metric
\beq
ds^2=-dt^2 + a^2(t)dx\cdot dx.
\eeq{4}
As usual, we denote the time derivative with an overdot, and we 
define the Hubble parameter as $H=\dot{a}/a$.
By decomposing a minimally-coupled, free scalar field $\phi$ of mass $m$
into plane waves of co-moving momentum $k$, we find the equation of motion
\beq
\left[{d\over dt} a^3(t) {d\over dt} + a(t)k^2 + a^3(t)m^2\right]\phi(t,k)=0.
\eeq{5}
The Feynman Green's function obeys the equation
\beq
\left[{d\over dt} a^3(t) {d\over dt} + a(t)k^2 + a^3(t)m^2\right]G_F(t,t',k)
=i\delta(t-t').
\eeq{6}
This function can be written in terms of two independent solutions to the 
homogeneous wave equation, $\phi^+(t,k)$ and $\phi^-(t,k)$, as:
\beq
G_F(t,t',k)= \phi^+(t,k)\phi^-(t',k)\theta(t-t') + 
\phi^-(t,k)\phi^+(t',k)\theta(t'-t).
\eeq{7}
Eq.~(\ref{6}) gives us the normalization of $\phi^\pm(t,k)$
\beq
\phi^-(t,k)\dot{\phi}^+(t,k)- 
\phi^+(t,k)\dot{\phi}^-(t,k)= {i\over a^3(t)}.
\eeq{8}
Next, we have to find the appropriate basis for $\phi^\pm(t,k)$. In the 
presence of a space-like boundary at some fixed time $t=t_0$, 
the action for the scalar field is
\beq
S_{bulk}+S_{boundary}=
{1\over 2}\int d^4x \sqrt{-g} 
\left[ g^{\mu\nu}\partial_\mu \phi \partial_\nu \phi +
m^2 \phi^2 \right] + {1\over 2}\int_{t=t_0} 
d^3x d^3y \sqrt{h(x)}\sqrt{h(y)} \phi(x) \kappa(x,y)\phi(y).
\eeq{9}
Here, $h_{ij}$ is the induced metric on the boundary, and
$\kappa(x,y)$ encodes the initial quantum state of the scalar~\cite{sspds}.
The Bunch-Davies vacuum corresponds to choosing a specific 
boundary action, which, among other properties, is invariant 
under space translations: $\kappa(x,y)=\tilde{\kappa}_{BD}(x-y)$. By 
performing a Fourier transform, it can be written somewhat formally as 
\beq
S_{boundary}={1\over 2} a^3(t_0)\int {d^3k \over (2\pi)^3} \left[
\kappa_{BD}(k)|\phi^+(t_0,k)|^2-\kappa_{BD}(k)|\phi^-(t_0,k)|^2 \right].
\eeq{10}
The precise form of $\kappa_{BD}(k)$ is immaterial to our computation.
What matters is that two modes $\phi^\pm(t,k)$ obey
\beq
\left.\left[-{d\over dt} \pm \kappa_{BD}\right]\phi^\pm(t,k)\right|_{t=t_0}=0.
\eeq{11}
Actually, Eq~(\ref{11}) is valid more generally. By considering a generic
translation-invariant boundary term $\kappa(x,y)=\tilde{\kappa}(x-y)$, and
extremizing  with respect to {\em all} variations $\delta\phi$ (including 
those that do not vanish at the boundary) we find the equation
\beq
\left.\left[-{d\over dt} \pm \kappa\right]\phi^\pm(t,k)\right|_{t=t_0}=0,
\eeq{11'}
where $\kappa(k)$ is of course the Fourier transform of $\tilde{\kappa}(x)$.

Suppose now that the boundary conditions are changed. To take into account 
the possibility that the change is different for the two modes, we
define $\kappa^{+}\equiv\kappa_{BD} +\delta\kappa^{+}$, 
$\kappa^{-}\equiv\kappa_{BD} +\delta\kappa^{-}$. 
To first order in the change, we have
\beq
\left.\left[-{d\over dt} \pm \kappa_{BD}\right]
\delta\phi^\pm(t,k) \pm \delta\kappa^{\pm}\phi^\pm(t,k)\right|_{t=t_0}=0.
\eeq{12}
Now, expand the modes $\delta\phi^{\pm}(t,k)$ to first order in the 
perturbation as 
\beq
\delta\phi^{+}(t,k) = \delta b^+ \phi^{-}(t,k), \qquad
\delta\phi^{-}(t,k) = \delta b^- \phi^{+}(t,k) .
\eeq{13}
Substituting this expansion into Eq.~(\ref{12}), we find
\beq
\delta b^{+}= {1\over 2} \delta \kappa^{+} {\phi^+(t_0,k) \over
\dot{\phi}^-(t_0,k)}, \qquad
\delta b^{-}= -{1\over 2} \delta \kappa^{-} {\phi^-(t_0,k) \over
\dot{\phi}^+(t_0,k)}.
\eeq{14}

Define next the symmetric Green's function 
$G^{(1)}(t,t')=(1/2)\langle 0| [\phi(t) \phi(t') + \phi(t')\phi(t)]|0\rangle$.
After using the normalization condition Eq.~(\ref{8}), and the identity
$\phi^- (t_0,k)\dot{\phi}^+ (t_0,k)=- \dot{\phi}^-(t_0,k)\phi^+(t_0,k)$, we
can express the change in $G^{(1)}(t,t')$ as
\beq
\delta G^{(1)}(t,t',k)= ia^3(t_0) \left[ \delta\kappa^+ \phi^{+\,2}(t_0,k)
\phi^- (t,k)\phi^- (t',k) + \delta\kappa^- \phi^{-\,2}(t_0,k)
\phi^+ (t,k)\phi^+ (t',k)\right].
\eeq{15}
$G^{(1)}$ is particularly useful when computing expectation values of composite
operators by point-spitting. In the next section we use it to
compute one such operator: the stress-energy tensor.
\section{Back-Reaction and Bounds}
We will consider just one example of IR irrelevant boundary perturbation
encoding the effects of high-energy physics. It is the dimension-four 
boundary term~\cite{sspds}
\beq
\delta S_{boundary} =a(t_0){\beta\over 2M}\int d^3x (\nabla \phi)^2.
\eeq{16}
It induces a change $\delta \kappa^{\pm}(k)=\pm \beta k^2/a^2(t_0)M$.
From now on, we shall choose for convenience $a(t_0)=1$. 

The change in boundary conditions induces a change in $G^{(1)}(t,t')$. 
This change is {\em finite} [see Eq.~(\ref{15})]
and {\em regularization independent} away from the boundary, i.e. as
long as $|t-t_0|>1/M$, $|t'-t_0|>1/M$~\footnote{To play safe, one should take 
an $M'$ somewhat smaller than $M$. This will not affect significantly our
estimates.}. Now, when we compute the
expectation value of the stress-energy tensor $T_{\mu\nu}=
\partial_\mu \phi\partial_\nu \phi - (1/2)g_{\mu\nu}(\partial_\lambda \phi
\partial^\lambda \phi + m^2\phi^2)$ by using $G^{(1)}(t,t')$, we get a
new contribution to first order in $\delta\kappa^\pm$. Notice that in our 
formalism one does not have to independently specify the vacuum state: 
that information is already contained in $G^{(1)}(t,t')$.

Again, the first-order contribution
is unambiguous and finite as long as $|t-t_0|>1/M$. 
The effect of the 
perturbation in, say, $\langle 0|T_\mu^\mu(t,x)|0\rangle$, 
is easy to estimate for times
$1/M < |t-t_0| \ll 1/M_0$, where $M_0\equiv \max{(H, m)}$. In this case, 
the VEV is determined by an integration over momenta $M>k\gg M_0$. For 
these momenta, much larger than either $H$ or $m$, we have ($\Delta t \equiv
|t-t_0|$): 
\beq
\phi^+(t,k)\phi^-(t_0,k) \approx {1\over 2|k| } e^{i|k|\Delta t} ,
\qquad
\phi^-(t,k)\phi^+(t_0,k) \approx {1\over 2|k| } e^{-i|k|\Delta t}, \qquad
1/M<\Delta t \ll 1/M_0. 
\eeq{17}
By using this estimate, we find
\beq
\delta \langle 0|T_\mu^\mu(t,x)|0 \rangle \approx
{\beta \over M}\int_{|k|>M_0}{d^3k \over (2\pi)^3} |k|^2 \sin(2|k|\Delta t)
\approx {3\over 16\pi^2}{\beta \over M}(\Delta t)^{-5}, \qquad
1/M<\Delta t \ll 1/M_0.
\eeq{18}

Equation~(\ref{18}) is our main estimate. Now it is clear what to do next: we 
substitute Eq.~(\ref{18}) into Einstein's equations for the background 
geometry and we ask ourself which bounds must be satisfied by the 
``new-physics'' parameter $\beta$ so that the back-reaction is negligible
at all times $\Delta t > 1/M$~\footnote{Clearly, at $\Delta t < 1/M$ our
estimate becomes ambiguous, i.e. regularization-dependent.}.

The first bound, Eq.~(\ref{1}) is the most general. It follows from demanding
that the change in the
vacuum energy, proportional to $ \delta \langle 0|T_\mu^\mu(t,x)|0 \rangle$, 
is much 
smaller than the unperturbed vacuum energy  $V=3H^2/8\pi G$. Setting
$\Delta t \sim 1/M$ we find 
\beq
\delta \langle 0|T_\mu^\mu(t_0+1/M,x)|0 \rangle=
{3\over 16\pi^2} \beta M^4 < {3H^2\over 4\pi G},
\eeq{19}
whence Eq.~(\ref{1}).
The bound is parametrically $H^2/M^2$, but it is multiplied by a coefficient
$2\pi/ GM^2$, which can be easily as large as $10^6$ if $M\sim 10^{16}\, 
\mbox{GeV}$.

The second bound comes from demanding that an inflationary period exists 
for some time after
$t_0$. This is reasonable  
whenever the scale of inflation, $H$, is smaller 
than the scale of new physics, $M$.
It requires the Hubble parameter to remain almost constant over a 
Hubble time, i.e. $\dot{H} \ll H^2$. Since Eq.~({\ref{18}) introduces a 
time dependence into $H$, we must set
\beq
|\delta \dot{H}| = {5\over 2H}{8\pi G\over 3} {3\over 16\pi^2} \beta M^5 
< | \dot{H}|
\eeq{20} 
By introducing the slow-roll parameter  
$\epsilon=|\dot{H}/H^2|$, 
we get Eq.~(\ref{2}). For the  choice of parameters 
$H=10^{14}\, \mbox{GeV}$, $M=10^{16}\, \mbox{GeV}$,
we can allow for a coefficient $\beta\sim \epsilon$, which may still 
fall within the range of future detectability.

The third bound requires a further assumption, namely that the 
quantum-corrected potential, inclusive of 
$\delta \langle 0| T_\mu^\mu |0 \rangle$, can be written as a function of a 
single scalar $\phi$ (the inflaton e.g.). 
If this is the case, then it is easy to show that, by introducing a
second slow-roll parameter, $\eta= |\ddot{\phi}/H\dot{\phi}|$, 
we have 
$2\epsilon\eta=|\ddot{H}/ H^3|$. So, from
\beq
|\delta\ddot{H}| < |\ddot{H}|=2\epsilon \eta H^3, 
\eeq{21}
we arrive at the last bound, Eq.~(\ref{3}).
This is the most stringent bound: by requiring only that both
$\eta$ and $\epsilon$ have ``standard,'' $O(10^{-1})$ values, we find 
$\beta \sim \epsilon\eta 10^{-2}\sim 10^{-4}$, 
easily below any chance of detection.
\subsection*{Acknowledgments}
We would like to thank G. Gabadadze and L. Randall for comments. 
M.P. is supported in part by NSF grants PHY-0245068 and PHY-0070787. 

\end{document}